\documentclass[%
 reprint,
 amsmath,amssymb,
 aps,
 onecolumn,
]{revtex4-1}
\usepackage{graphicx}
\usepackage{dcolumn}
\usepackage{subfigure}
\usepackage{bm}
\usepackage[utf8]{inputenc}
\usepackage{mathrsfs}
\usepackage[T1]{fontenc}
\usepackage[mathlines]{lineno}
\usepackage{mathrsfs}
\usepackage{multirow}
\usepackage{amsmath}
\usepackage{amssymb}
\usepackage{graphicx}
\usepackage{subfigure}
\usepackage{esint}
\usepackage{gensymb}
\usepackage{float}
\usepackage{lineno}
\usepackage{setspace}

\providecommand{\tabularnewline}{\\}
\makeatother
\usepackage{babel}

\begin{document}
\title{Research progress on advanced positron acceleration}
\begin{spacing}{2.0}
\author{Meiyu Si}
\affiliation{Institute of High Energy Physics, Chinese Academy of Sciences, Beijing 100049, China}
\affiliation{University of Chinese Academy of Sciences, Beijing 100049, China}
\author{Yongsheng Huang} 
\email{ huangysh59@mail.sysu.edu.cn}
\affiliation{School of Science, Shenzhen Campus of Sun Yat-sen University, Shenzhen 518107, China}
\affiliation{Institute of High Energy Physics, Chinese Academy of Sciences, Beijing 100049, China}

\begin{abstract}
Plasma wakefield acceleration (PWFA) is a promising method for reducing the scale and cost of future electron-positron collider experiments by using shorter plasma sections to enhance beam energy. While electron acceleration has already achieved breakthroughs in theory and experimentation, generating high-quality positron beams in plasma presents greater challenges, such as controlling emittance and energy spread, improving energy conversion efficiency, and generating positron sources. In this paper, we have summarized the research progress on advanced positron acceleration schemes, including particle beam-driven wakefield acceleration, laser-driven wakefield acceleration, radiation-based acceleration, hollow plasma channels, among others. The strengths and weaknesses of these approaches are analyzed, and the future outlook is discussed to drive experimental advancements.
\end{abstract}
\maketitle

\section{Introduction}
Traditional accelerators are limited by the material breakdown threshold, with the acceleration gradient typically falls within the range of a few $\mathrm{MeV/m}$ to several tens of $\mathrm{MeV/m}$ \cite{Simakov,Lilje}. The lower gradients require longer accelerator structures to enable particles to reach higher energies, thereby increasing the scale and cost of experiments. Plasma wakefield acceleration (PWFA), powered by the nonlinear ponderomotive force of lasers or an ultrarelativistic charged particle beam, generate an acceleration gradient of $\mathrm{GV/m}$ that allows trapped electrons in the wakefield to be accelerated to high energies\cite{Tajima}.  Therefore, plasma-based charged particle accelerators\cite{Gschwendtner} are an attractive alternative to traditional circular and linear accelerators, offering the potential for higher acceleration rates and a more compact facility\cite{Dunne,Bingham}. High-energy charged particle beams, such as electrons and positrons, are obtained for a variety of applications, including fundamental research\cite{Griffiths,Guessoum}, materials science\cite{Marchut}, medical imaging\cite{Ciarmiello}, radiation therapy\cite{Hogstrom}, they can be used for the construction of next-generation large particle colliders, such as the International Linear Collider (ILC)\cite{ILC}, the Compact Linear Collider (CLIC) \cite{clic}, the Circular Electron Positron Collider (CEPC)\cite{cepc}, to explore the properties of fundamental particles and discover new particles.  

Particle-driven plasma wakefield acceleration is being pursued in large-scale projects, including CERN's Advanced Proton Driven Plasma Wakefield Acceleration Experiment (AWAKE)\cite{Bracco} , SLAC's Facilities for Accelerator science and Experimental Test (FACET)\cite{Hogan}, the Berkeley Lab Laser Accelerator (BELLA) Laser-Plasma Accelerator (LPA) experiment\cite{Leemans}. Significant progress has been made in the research of plasma-based accelerators for generating high-energy electron beams\cite{Mangles,Leemans-W-P,Faure}. A laser pulse with a peak power of 850 $\mathrm{TW}$ generates an electron beam, which carries several hundred picocoulombs of charge and multiple quasi-monoenergetic peaks. The highest peak energy reaches 7.8 $\mathrm{GeV}$ with the beam divergence of 0.2 $\mathrm{mrad}$\cite{Gonsalves}. In order to achieve a collider, both gradient acceleration of positrons and gradient acceleration of electrons are equally important. However, achieving high-quality positron beam acceleration in a plasma is more challenging. This is due to the narrow nonlinear bubble region formed by plasma electrons for positrons, the variation of the acceleration field along the axial direction, which increases energy spread, and the transverse field defocusing, leading to increased emittance\cite{Lotov}. The lack of suitable ultra-relativistic positron sources has resulted in fewer experimental studies on wakefield acceleration\cite{Leemans}. Therefore, research on generating high-quality ultra-relativistic positron sources\cite{Alejo,Chen} and innovations in plasma-based acceleration schemes are both crucial. We provide a summary of positron acceleration schemes, including particle beam-driven wakefield acceleration, laser-driven wakefield acceleration, radiation-based acceleration, hollow plasma channels, among others.

A relativistic electron beam\cite{Diederichs} moving through plasma forms bubble structures, with the favorable region for positron acceleration located between the first and second bubbles. This region is extremely narrow (acceleration region: $\partial |E_z|/\partial z <0$), resulting in an energy spread of about several percent, which imposes a high level of precision in the positron injection position\cite{Lotov}. Reference \cite{Wang} proposed a new method that involves two electron beams entering a system composed of a plasma and an embedded thin foil. Approximately $10^7$ positrons can gain energy $\sim$ 5 $\mathrm{GeV}$ over 1 $\mathrm{m}$, with the relative energy spread of 16$\%$ and the normalized emittances of 40 $\mathrm{mm}$ $\mathrm{mrad}$. However, the issue of a narrow acceleration region still exists. For a positron driver\cite{Lee,Blue,Corde}, plasma electrons are attracted towards the path rather than being pushed away, the main body of the positron beam to lose energy while exciting this wakefield. Positrons behind in the same beam can then be accelerated by the same wakefield when they change sign, the efficiency of energy transfer gradually decreases from front to back. The acceleration field is weaker of about tens of $\mathrm{MeV/m}$ compared to that driven by an electron beam. The focusing field is nonlinear in the transverse coordinate system and varies along the axis of the beam, resulting in an increase in the emittance of the witness beam. 

Laser driving in the plasma can generate higher acceleration gradients, with the order of acceleration field $E_0(V/m)\approx96\sqrt{n_0(cm^{-3})}$, where $n_0$ is the density of plasma electrons\cite{L-L-Yu}. Laguerre-Gaussian (LG) laser pulse with high intensity has been proposed for accelerating positrons in a uniform plasma, which can generate a thin plasma electron filament along the axis to excite high-gradient positron accelerating field and focusing wakefields in the nonlinear regime\cite{Vieira}. Hollow electron beam\cite{Thales} drive is equally effective, with the drawback being the acquisition of the electron beam source. According to the parameters of FACET, the Neeraj Jain \cite{Jain} utilized a hollow electron beam as the driver beam to simulate the generation of a similar nonlinear distribution field. The acceleration gradient reached 10 $\mathrm{GV/m}$, resulting in the acceleration of a 23 $\mathrm{GeV}$ positron beam to 35.4 $\mathrm{GeV}$, with a maximum energy spread of 0.4$\%$. However, their energy transfer efficiency is very limited and may tend to exhibit dynamic instabilities. A hollow plasma channel\cite{Kimura,Silva} has been proposed as a potential candidate for plasma acceleration and can provide radial confinement for positrons. Reference \cite{Yi} proposes using a 2 $\mathrm{TeV}$ proton beam to drive the acceleration of 1 $\mathrm{GeV}$ positrons within a cylindrical vacuum plasma channel (with a channel density of $10^{15} cm^{-3}$), achieving a peak positron energy of up to 1.6 $\mathrm{TeV}$ within 1.1 kilometers. To overcome the natural Coulomb dispersion, quadrupole magnets are added outside the channel to enhance the focusing. In Reference \cite{Wei}, an asymmetric electron beam is used to create a stable accelerating and focusing field for positrons within a hollow plasma channel (with channel density of $3.11\times10^{16} cm^{-3} $). The conversion efficiency can reach 30$\%$, and energy control is maintained at a 1$\%$ level. Based on the above methods, it may be necessary to consider how to obtain a hollow-driven electron beam source and the ease of preparing channels for positron acceleration using hollow channels.

At the same time, radiation-based positron acceleration schemes have also been under exploration. In Reference \cite{Xu,Yan-Ting,Terzani}, an all-optical approach is proposed, which includes the generation, injection, and long-distance acceleration of positrons to achieve GeV-level energies. The acceleration field is generated by coherent transition radiation (CTR) when an electron beam passes through the high-density target at the vacuum interface and is constrained by a magnetic field in the propagation direction. For longer-distance acceleration, in a recent publication, reference \cite{Xu2} replaced an external magnetic field with a plasma channel, achieving the acceleration of 1 $\mathrm{GeV}$ positrons to 128 $\mathrm{GeV}$ over a distance of 1 meter. In Reference \cite{Si}, an analysis is conducted on the longitudinal radiation acceleration field generated by the oscillations of a surface electron film driven by a relativistic electron beam in a dense plasma micro-tube.  The field strength is linearly proportional to the charge of the electron beam. In this radiation field, external injection of positrons can achieve an energy acceleration of 1 $\mathrm{GeV}$ within 4.2 centimeters\cite{Si2}, and it is also possible to consider cascade acceleration and series acceleration of multiple positron beams. The efficiency of energy transfer from the electron beam to one positron bunch or three positron bunches simultaneously, could reach up to 20$\%$ and 40$\%$. In Reference \cite{Jie-zhao,Zhao}, it is proposed to drive positron acceleration using strong coherent terahertz radiation generated when an injected electron ring beam passes through one or multiple solid targets. The acceleration gradient reaches up to 2.5 $\mathrm{TeV/m}$, with a relative energy spread of $\Delta E/E = 2\%$. These methods provide new possibilities for positron acceleration and application fields. Currently, most of the proposed acceleration schemes are based on simulation analysis. To make them experimentally feasible, further considerations need to be taken into account, including the generation of positron sources, injection methods, and the fabrication of hollow channel acceleration structures.

Figure 1 illustrates some effects related to positron acceleration, primarily based on particle-in-cell (PIC) simulation analysis, including their longitudinal acceleration gradients and final energy spread. The longitudinal acceleration fields are mainly concentrated in the range of 0.1 $\mathrm{GV/m}$ to 50 $\mathrm{GV/m}$, with control of the positron energy spread kept below 5$\%$. Achieving ultra-high acceleration gradients, such as hundreds of $\mathrm{GV/m}$ or even $\mathrm{TV/m}$, requires compact charged particle beam driving with high charge densities or strong laser field driving. And $\eta$ is refers to the energy conversion efficiency from driving beam to positrons. 
Table 1 shows several representative positron acceleration methods, including acceleration mechanisms, the charge and energy of driving and witness beams, and acceleration results of the positron beam. 
The main text of this paper is divided into four sections for discussion. In Sec. A, the advantages of plasma accelerators compared to traditional accelerators are analyzed, focusing on the acceleration of charged particle beams and the generation of radiation sources. In Sec. B, challenges associated with positron beam acceleration based on plasmas are analyzed. In Sec. C, representative proposed acceleration schemes are examined. In Sec. D, experimental progress in plasma-based acceleration is discussed, including the acceleration of electrons and positrons.

\begin{figure}
	\centering
	\includegraphics[width=0.8\textwidth]{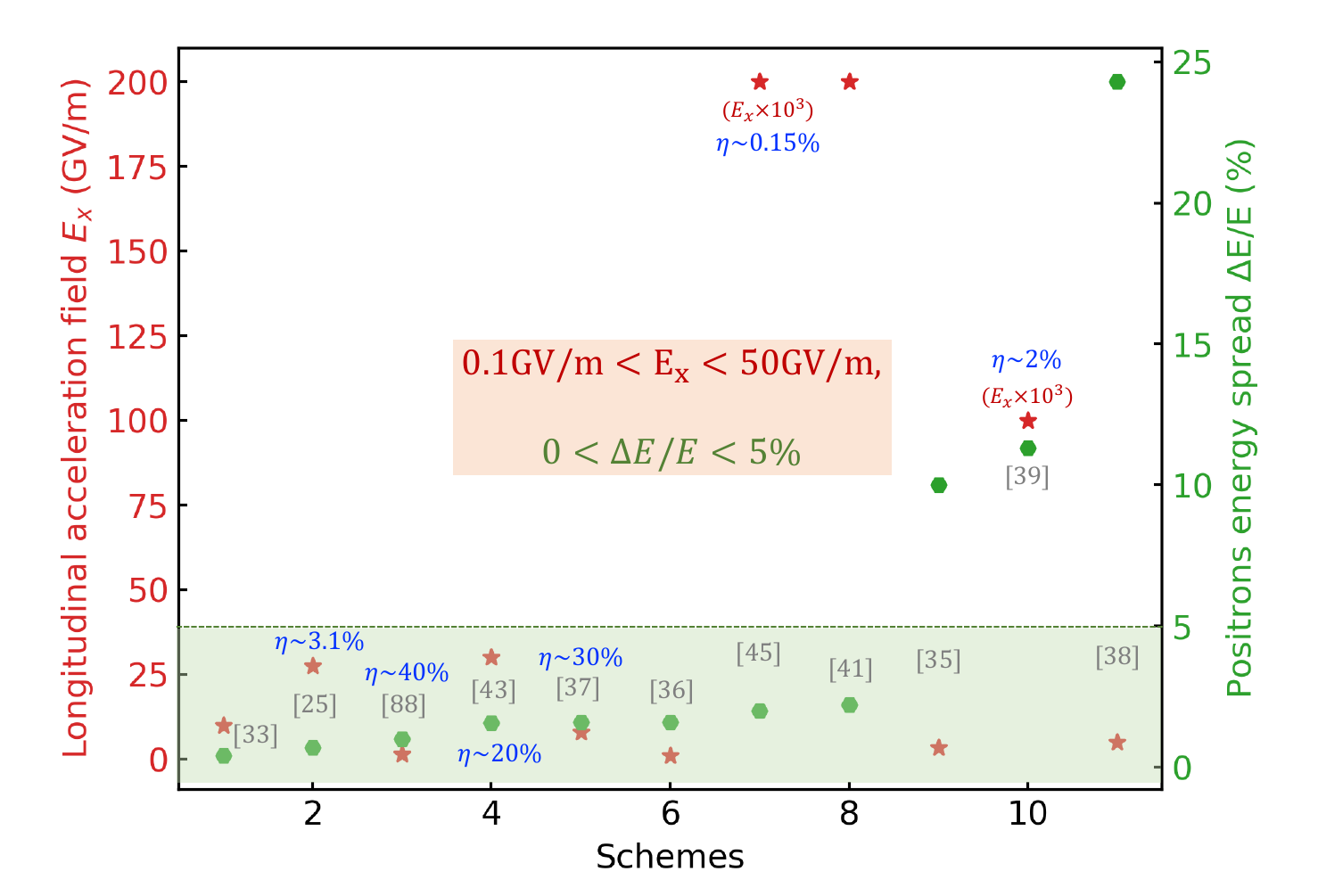}
	\caption{({\bf a}) Figure 1 illustrates some effects related to positron acceleration, primarily based on particle-in-cell (PIC) simulation analysis, including their longitudinal acceleration gradients and final energy spread. The longitudinal acceleration fields are mainly concentrated in the range of 0.1 $\mathrm{GV/m}$ to 50 $\mathrm{GV/m}$, with control of the positron energy spread kept below 5$\%$. }
	\label{Figure1}
\end{figure}

\begin{table}[h]
\centering
\begin{tabular}{p{2cm}p{5cm}p{3.5cm}p{3.5cm}p{4cm}}
\hline
References & Acceleration Mechanism & Driving Beams & Witness Beams & Results \tabularnewline
\hline
\cite{Wei} & $e^-$ and plasma tube\newline $(3.11\times 10^{16} cm^{-3})$ & 2 nC/5.11GeV $e^-$\newline asymmetric beam & 0.64nC/10.2GeV $e^+$  & 0.49nC/14.6GeV $e^+$ in 0.9m,\newline RMS energy spread of 1.6$\%$
\tabularnewline
\cite{Yan-Ting} & Collision of a twisted Gauss \newline laser in NCD plasma, \newline phase-locked-acceleration& Driving laser $a_0=265$ to generate GeV $e^-$ & NCS/BW process($e^+$)  & 55pC/1GeV, sub-fs $e^+$
\tabularnewline
\cite{Xu2} & CTR and electron beam &  26.6nC/45GeV $e^-$  & 290pC/1GeV $e^+$ & 10pC/126.8GeV $e^+$ in 1m, RMS energy spread of 2.2$\%$
\tabularnewline
\cite{Si2} & Relativistic electron beam,\newline dense plasma micro-tube & 3.402nC/1GeV $e^-$ &111pC/1GeV $e^+$  &90pC/2GeV $e^+$ in 4cm, RMS energy spread of 1.56$\%$

\tabularnewline
\cite{Zhao} & $e^{-}$ ring and THz radiation,\newline one or multiple solid targets & 0.1$n_c$/2GeV $e^-$   &100MeV  & $\delta_E$=200MeV, RMS energy spread of 2$\%$
\tabularnewline
\hline
\end{tabular}
\caption{Table 1 shows several representative positron acceleration methods, including acceleration mechanisms, driving beam parameters, witness positron beam parameters, and acceleration results. NCD: near critical density. NCS: nonlinear Compton scattering. 
BW: Breit-Wheeler process. $n_c=1.2\times 10^{27} m^{-3}$.}
\end{table}

\subsection{The advantages of plasma accelerators}
Accelerators\cite{Chernyaev,Conard} play an irreplaceable role in advancing scientific research\cite{Bhandari,offe}, medical treatment\cite{Amaldi,Wieszczycka,Cline,Sidky}, and industrial progress\cite{Selim}. With the continuous development of science, there is a growing demand for higher-energy accelerators\cite{Gourlay}. Higher energy allows the study of the properties of matter at smaller scales, aiding in the validation of existing theories, development of new theoretical models, and exploration of topics such as dark matter and dark energy. Since the accelerating field inside traditional superconducting radiofrequency cavities is approximately tens of  $\mathrm{MV/m}$, in order to achieve high energy gain, it is necessary to increase the length of the accelerator, leading to increased accelerator size and costs. For example, the circumference of the larger collider reaches several tens of kilometers\cite{ILC,clic,cepc,lhc}. Therefore,  the use of plasma accelerators\cite{Nakajima,Hooker,Rosenzweig} is proposed, which eliminates the limitations associated with material ionization thresholds and can support and sustain extreme electric fields\cite{Malka}. In 1979,  the theoretical work of  Tajima and  Dawson\cite{Tajima} analyzed the nonlinear ponderomotive force of intense laser pulses, which can excite plasma oscillations, generate a wakefield, and achieve electric field strengths of up to hundreds of $\mathrm{GV/m}$ and high energy efficiency. As shown in Figure 2(a), the accelerating field in traditional accelerators is less than 100 $\mathrm{MV/m}$, while the accelerating gradient in plasma-based accelerators can exceed 100 $\mathrm{GV/m}$. This implies that obtaining a 100 $\mathrm{MeV}$ energy gain for a charged particle beam in traditional accelerators requires an acceleration distance on the order of meters, whereas for the new principle accelerator, achieving the same energy gain only requires an acceleration distance on the order of millimeters. A large acceleration gradient can accelerate injected charged particles\cite{Faure,Litos(2014)}, enabling the construction of more compact accelerators\cite{Kurz}. The laser beam can also be replaced with a relativistic charged particle beam\cite{Huang,Lu} to drive the wakefield bubble, such as an electron beam, positron beam, or proton beam.

The plasma accelerator also can be used to produce high-brightness beam of $\mathrm{keV}$ and $\mathrm{MeV}$ photons by betatron radiation\cite{Schwoerer,Schlenvoigt}. Relying on the oscillation of relativistic electrons generated in laser plasma accelerators is the most compact and promising scheme, such as plasma wiggler\cite{Rousse}, laser wiggler\cite{Hartemann}, and X-ray free-electron lasers\cite{Schlenvoigt,Gruener}. The schematic diagram of radiation generation by plasma wiggler, as shown in Figure 2(b), illustrates the transmission of laser fields in a plasma background, forming a bubble structure. This structure facilitates the acceleration of electrons to relativistic energies. Therefore, the oscillation of electrons within the cavity leads to radiation emission along the direction of electron momentum. The peak energy of radiation can be obtained by $E(eV)=1.45\times 10^{-21}\gamma ^2n_e(cm^{-3})r_0(\mu m)$, where $\gamma$ is the Lorentz factor, $n_e$ is the density of plasma electrons, and $r_0$ is the amplitude of the oscillation motion\cite{Malka(2008),Kostyukov}. 
High-power terahertz radiation (<10 $\mathrm{THz}$) was observed to be emitted from laser-wakefield accelerated electrons with 4 $\mathrm{mJ}$ energy and coversion efficiency of ~0.15$\%$\cite{Pak,Hooker}.  

\begin{figure}
	\centering
	\includegraphics[width=0.65\textwidth]{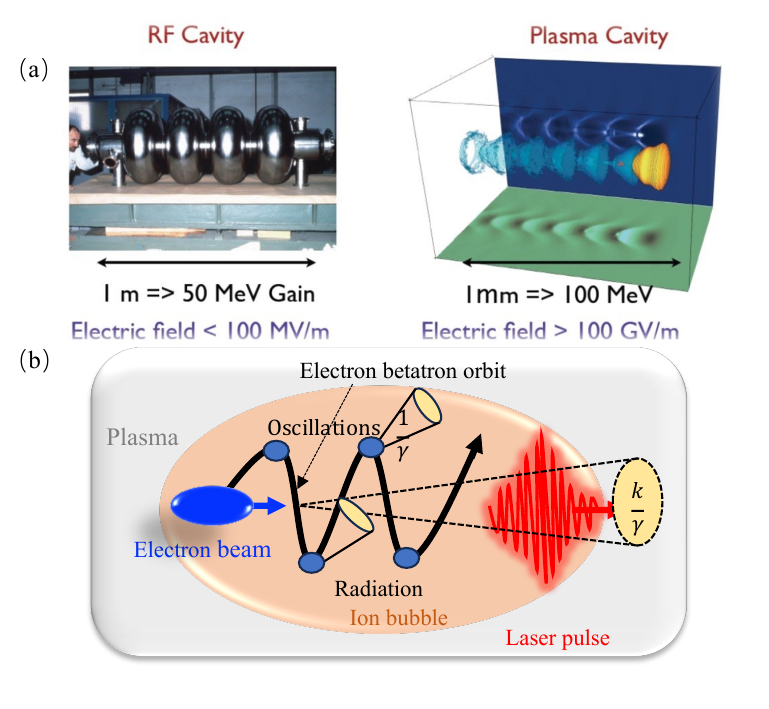}
	\caption{({\bf a}) The differences between traditional radiofrequency cavity accelerators and plasma cavity accelerators,  such as the most critical parameter, the acceleration gradient. From Malka, V.\cite{Malka}. ({\bf b}) The schematic diagram of radiation generation, illustrates the transmission of laser fields in a plasma background, forming a bubble structure. The oscillation of electrons within the cavity leads to radiation emission along the direction of electron momentum. }
	\label{Figure2}
\end{figure}

\subsection{The challenges of positron beam acceleration}
As plasma accelerators offer high-gradient advantages, they enable charged particles such as electrons, positrons, protons, carbon ions, and others to attain high energies over shorter acceleration distances. Positrons, as the antimatter counterparts of electrons, have a wide range of applications. They are particularly crucial in future collider experiments. However, the acceleration of positron beams in plasmas still poses long-standing fundamental challenges. Firstly, the difficulty in generating positron sources in experiment. Positrons are typically generated in the decay of radioactive isotopes, such as $^{22}\text{Na}$ and $^{64}\text{Cu}$, exhibiting a broad energy spectrum and proving challenging to capture and accelerate. Additionally, the intensity provided by radioactive sources is in the range of  $10^6-10^8$ $e^+/s$ in the forward $2\pi$ radians, which is 4 to 6 orders of magnitude lower than the intensity usually required for high-energy physics applications\cite{Hugenschmidt,Chaikovska,Golge}. Therefore, the proposal involves generating positrons through the interaction of lasers with high-energy electron beams. This process typically includes nonlinear Compton scattering and the Bethe-Heitler (BH) process\cite{Y-H-Chen-2018}. The intensity required for a positron source is several orders of magnitude higher than existing facilities. For the FCC-ee, a positron beam intensity of $2.1\times10^{10}$ particles (3.35 nC) is required during injection\cite{FCC}.

The second challenge lies in the precision control of positron injection and the uniformity of the acceleration field gradient of plasma accelerator. 
Due to the very short effective region for positron acceleration within the bubble structure, high-precision control of the phase is required for the external injection of positrons. Therefore, some all-optical methods have been proposed to address the issue of positron injection, integrating the generation source with the acceleration process\cite{Xu,Yan-Ting}. The control over the maintenance time of the acceleration gradient is crucial to achieve uniform acceleration and obtain high energy. Then, in the process of positron acceleration, it is necessary to effectively control its energy spread to achieve stable and efficient acceleration results, which can lead to the instability and loss of the accelerated beam. Compared to bubble acceleration, hollow plasma shows better results, but the fabrication method and precision of this structure are also challenges that need to be addressed\cite{Yi}.

\subsection{Positron beam acceleration schemes}
\subsubsection{Plasma wakefield acceleration of positron beam}
Several positron acceleration schemes based on plasma wakefield acceleration have been proposed. Firstly, there is the wakefield driven by an electron beam, which forms a bubble in the plasma. Unlike electrons, whose acceleration phase occurs within the first bubble, allowing for effective longitudinal acceleration and transverse focusing, the acceleration phase for positrons is located between the first and second bubbles, with the schematic diagram of positron acceleration is shown in Figure 3(a). Considering the energy spread, the effectiveness of positron acceleration is related to the slope of the longitudinal field $E_z$. For a positron beam, the slope of the acceleration field is $\partial |E_z|/\partial z >0$ in the corresponding phase. The bunch is continuously stretched in the longitudinal direction, leading to an increase in energy spread. For $\partial |E_z|/\partial z < 0$, the positron bunch acceleration is effective, but the longitudinal scale is very short like in Figure 3(b). Two electron beams are used here: one beam is employed to generate the wakefield forming the acceleration structure, and the second bunch passes through a plasma target to produce the positron witness bunch, with its corresponding phase in $\partial |E_z|/\partial z < 0$. Results from three-dimensional Particle-in-Cell (PIC) simulations show that positrons gain an energy boost of approximately 5 $\mathrm{GeV}$ within 1 meter of plasma\cite{Wang}.  There have been publications related to positron-driven wakefield acceleration in both theory, simulations and experiments. The basic principle is that the main positron beam excites the wakefield in the plasma, losing energy in the propagation process, while the back of the same beam, acting as the witness beam, gets accelerated, like in Figure 3(c)\cite{Lee}.  In the case of positron-driven, plasma electrons are attracted towards the central axis instead of being pushed away. As a result, the acceleration field gradient is lower than that of electron beam-driven, typically ranging from tens to hundreds of $\mathrm{MV/m}$. Simultaneously, due to the nonlinear focusing acting on longitudinal slices, the emittance of the positron beam increases. Proposing the transmission of positron bunches through a hollow plasma channel (ionized by an intense laser field) to enhance the acceleration gradient. Utilizing a hollow plasma channel can generate a tail field with a significant longitudinal component and a small transverse component, including the driving of electron bunch, positron bunch or proton\cite{Silva,Yi,Wei,Corde}.
 
\begin{figure}
	\centering
	\includegraphics[width=0.7\textwidth]{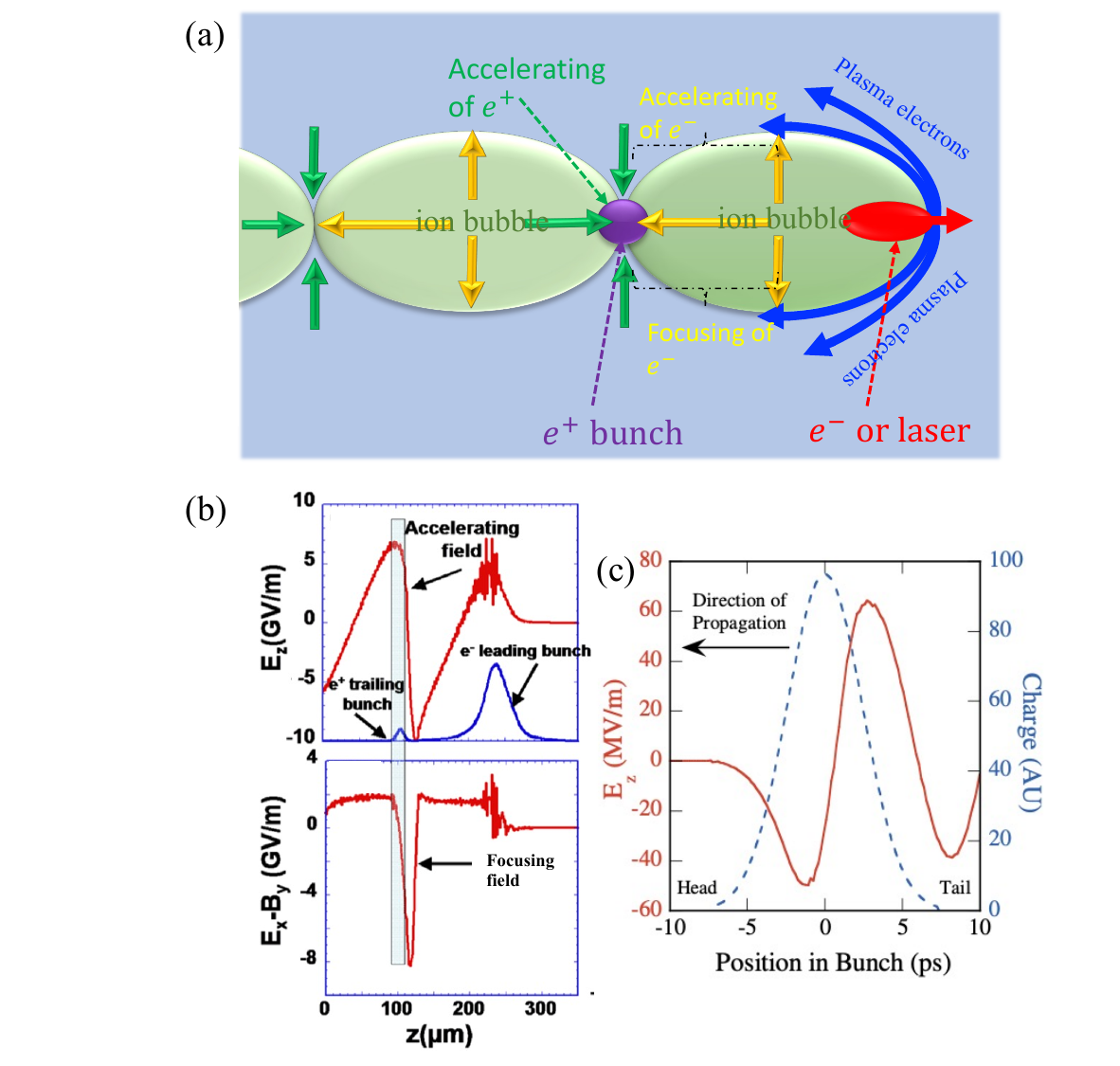}
	\caption{({\bf a}) shows the schematic diagram of the positron acceleration phase, with either electron-driven or laser-driven wakefields. Compared to the longitudinal acceleration and transverse focusing of electrons, the effective region for positrons is very short. ({\bf b}) shows the longitudinal wakefield ($E_z$) and the transverse wakefield ($E_x-B_y$) lineout along the z axis and x-z plane. The rectangle shows the positron phase with acceleration and focusing. From X. Wang et al\cite{Wang}. ({\bf c}) shows the longitudinal electric field in a plasma by a positron bunch. The head of the positron beam is decelerated, with a peak electric field of 50 $\mathrm{MV/m}$, while the trailing part of the same bunch is accelerated, with a peak electric field of 64 $\mathrm{MV/m}$. From B.E. Blue et al\cite{Blue}. }
	\label{Figure4}
\end{figure}

\subsubsection{Radiation-based acceleration of positron beam}
Radiation acceleration for electrons has been achieved experimentally. For example, Terahertz (THz) pulse-driven particle accelerators hold the potential for unprecedented control over the energy-time-phase space of particle beams. The linear acceleration of electrons achieved by the resulting terahertz pulse can obtain an energy gain of several thousand electron volts\cite{Nanni2015,Hibberd2020,Xu2021,Yu2023}. 
Due to the lack of externally injected positron sources, an all-optical approach has been proposed, integrating the generation, injection, and acceleration of positrons. Refences \cite{Xu} proposes that ultra-short electron bunches generated through laser wakefield acceleration pass through a solid target, producing a positron beam with a charge of approximately 12 $\mathrm{pC}$ via the Bethe-Heitler (BH) process. The electron bunch passes through the metal-vacuum interface, generating coherent transition radiation (CTR) to capture and accelerate the produced positron beam. The energy spectrum exhibits a cutoff energy of up to 1.5 $\mathrm{GeV}$. Martinez\cite{Martinez} proposed a head-on collision between an ultra-intense laser and a relativistic electron beam to generate positrons through nonlinear Compton scattering and Breit-Wheeler processes. The generated positrons have a charge of approximately 17 femtocoulombs, the energy can reach the GeV level and the emittance approximately 50 $\mathrm{mrad}$.  In reference \cite{Yu}, a short-period LG laser with an intensity of $1.2\times10^{21}$ $\mathrm{W/cm^2}$ is employed to effective capture, collimation, and longitudinal field acceleration up to 450 $\mathrm{MeV}$, with the divergence angle of 1 $\mathrm{^\circ}$ and the pulse width of 0.5 $\mathrm{fs}$. Positron acceleration schemes are continuously being innovated, such as the radiation acceleration. In addition to terahertz radiation, there is also the recently proposed mid-infrared radiation. The self-modulation of ultra-high-power relativistic laser pulses generates intense mid-infrared radiation pulses\cite{Gordon,Nie,Zhu,Nie2}, or a stable periodic structure of strong radiation acceleration fields produced by relativistic electron beams in high-density plasma micro-tube (such as metal tube) in Figure4(a)\cite{Si}. In this radiation field, external injection of positrons can achieve an energy acceleration of 1 $\mathrm{GeV}$ within 4.2 centimeters\cite{Si2}, and it is also possible to consider cascade acceleration and series acceleration of multiple positron beams. The efficiency of energy transfer from the electron beam to one positron bunch or three positron bunches simultaneously, could reach up to 20$\%$ and 40$\%$. Increasing the charge of the driving electron beam can enhance the intensity of the acceleration field, thereby enabling the generation of positrons with energies reaching hundreds of $\mathrm{GeV}$\cite{Xu2}. In Reference \cite{Zhao}, it is proposed to drive positron acceleration using strong coherent terahertz radiation generated when an injected electron ring beam passes through one or multiple solid targets in Figure 4(b). The acceleration gradient reaches up to 2.5 $\mathrm{TeV/m}$, with a relative energy spread of $\Delta E/E = 2\%$. This beam generates a thin plasma electron filament along the axis, simultaneously focusing and accelerating positrons. However, the energy transfer efficiency is limited, and there are dynamic instabilities.
 \begin{figure}
	\centering
	\includegraphics[width=\columnwidth]{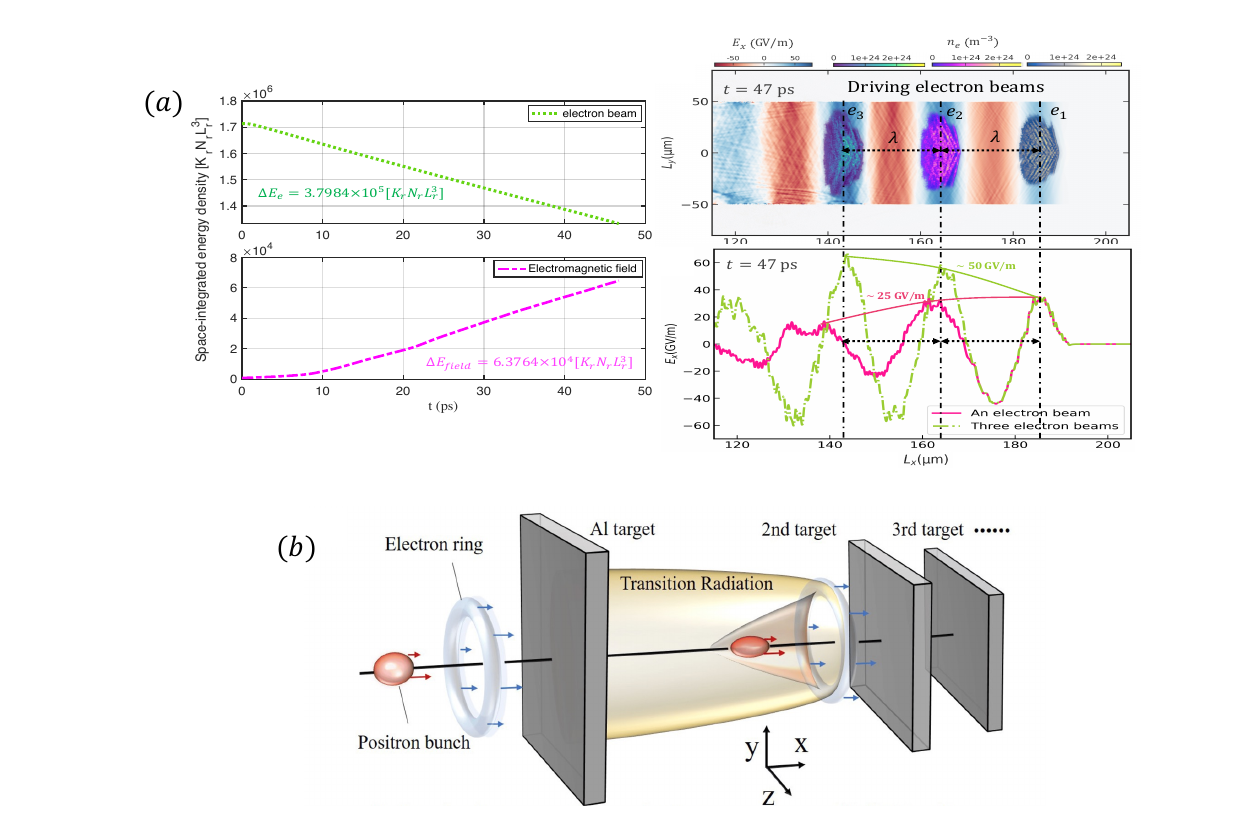}
	\caption{({\bf a}) shows the energy transfer efficiency and the radiation field distribution for a single electron beam and
three electron beams in a dense plasma micro-tube. The wavelength of the radiation field is in the mid-infrared range and is adjustable. From Meiyu Si et al\cite{Si}. ({\bf b}) shows the schematic diagram of the multistage THz driven positron acceleration by an electron ring. From Jie Zhao et al\cite{Zhao}.}
	\label{Figure4}
\end{figure}

\subsection{Experimental}
Plasma-based accelerators have the capability to generate ultra-short, high-charge, high-current, low-emittance, and narrow energy spread beams. In comparison to traditional accelerators, they can efficiently accelerate electrons, positrons, protons, and ions over short distances. The formation of wakefield structures in a plasma can be achieved through laser-driven and particle-beam-driven methods, and relevant experiments are currently underway. The relevant parameters of the provided drive beam in the experiment is show in Table II. For example, the early PWFA experiments were carried out at Stanford Linear Accelerator Center (SLAC) electron beams about 3$\mathrm{nC}$ with high energy (28.5 $\mathrm{GeV}$ and 42 $\mathrm{GeV}$) and the emittance less than $5\times 10^{-5} \mathrm mrad$, the bunch length $<20 \mathrm{\mu m}$ of ultra-short bunch and the long bunch about 730 $\mathrm{\mu m}$\cite{Joshi(2018)}. This experiment demonstrates that a 42 $\mathrm{GeV}$ drive electron beam generates an acceleration field of approximately 52 $\mathrm{GV/m}$ in a plasma with a density of $2.7\times 10^{17}$ $\mathrm{cm}^{-3}$, allowing the witnessed electron beam probing achieving a maximum energy of 85 $\mathrm{GeV}$\cite{Blumenfeld}. Compared to traditional accelerators, the plasma accelerator achieves the same energy gain in less than one meter, as opposed to the 3-kilometer length of a conventional accelerator. FACET provides two electron beams structure with an energy of 20.35 $\mathrm{GeV}$: One beam serves as the drive beam, generating wakefield structures in the plasma, the other beam acts as the witness beam undergoing acceleration. The results indicate that the electron beam, driven in a plasma with a density of $5\times 10^{16}$ $\mathrm{cm^{-3}}$ , generates a wakefield with an acceleration gradient of 6.9 $\mathrm{GV/m}$, enabling the witness beam to achieve a 9 $\mathrm{GeV}$ energy gain within a distance of 1.3 meters, with a mean energy spread of 5.1$\%$\cite{Litos}. The AWAKA is the first plasma wakefield acceleration experiment driven by proton beam with energy of 400 $\mathrm{GeV}$, allowing the witness electron bunch to accelerate from an initial energy of 10-20 $\mathrm{MeV}$ to GeV scale within a few meters\cite{Bracco}. Using a relativistically intense laser pulse with a peak power of 0.85$\mathrm{PW}$  and a normalized laser intensity of $a_0=8.5\times 10^{-10} \lambda[\mu m] \sqrt{I_0[W/cn]^2}=2.2$, it is possible to accelerate an electron beam to 8 $\mathrm{GeV}$ with a charge of 7.8 $\mathrm{pC}$\cite{Gonsalves,Steinke}. 

The 28.5$\mathrm{GeV}$ positron beam at SLAC, consisting of approximately $1.2\times 10^{10}$ particles with a pulse width of 2.4 $\mathrm{ps}$, passes through a 1.4 $\mathrm{m}$ plasma with a density of $1.8\times 10^{14}$ $\mathrm{cm^{-3}}$. The main body of the beam decelerates at a rate of 49 $\mathrm{MeV/m}$, while about $5\times 10^{8}$ positrons in the back of the same bunch gain energy at an average rate of approximately $\mathrm{MeV/m}$\cite{Blue}. The measured energy gain of positrons is approximately 79 $\mathrm{MeV}$,  but the acceleration gradient is lower than that of the plasma wakefield for electrons. The maximum energy gain was about 5 $\mathrm{GeV}$ with the energy spread can be as low as 1.8$\%$, the energy of driving positron bunches is 20.35 $\mathrm{GeV}$, the charge is about 2.2 $\mathrm{nC}$\cite{Corde}.  Two positron bunches are injected into the plasma, a drive bunch with the energy of 20.55 $\mathrm{GeV}$ and the charge of 480 $\mathrm{pC}$, a trailing bunch with the energy of 20.05 $\mathrm{GeV}$ and the charge of 260 $\mathrm{pC}$ about 100 $\mathrm{\mu m}$ behind. The acceleration gradient is 1.12 $\mathrm{GeV/m}$, enabling positrons to achieve an energy gain of 1.45 $\mathrm{GeV}$ within a plasma length of 1.3 meters and the charge of 85 $\mathrm{pC}$, and the acceleration results of the two positron bunches is show in Figure 5(a). Additionally, the conversion efficiency between the reduced energy in the drive bunch (below 19.9 $\mathrm{GeV}$) and the energy gained by the trailing bunch (~20.8 $\mathrm{GeV}$) is 40 $\%$\cite{Doche}. A hollow plasma channel accelerator for positron-driven experiments was conducted at FACET, forming a hollow channel by ionizing lithium vapor with number density of $8\times10^{16}$ $\mathrm{cm^{-3}}$ through a high-intensity laser field (about 34 $\mathrm{mJ}$ and 100 $\mathrm{fs}$ FWHM). The positron bunch has a mean energy of 20.35 $\mathrm{GeV}$ with the gradient about 230 $\mathrm{MeV/m}$\cite{Gessner,Linds}. The experimental setup is show in Figure 5(b), including the laser ionization to form a hollow plasma channel, as well as the measurement of the field size, the ({\bf b1}) shows the size of the acceleration field and the positions of the drive beam and witness positron beam. In reference \cite{Yonghong}, an experiment was conducted on the XingGuangIII laser facility to investigate positron acceleration by the sheath field on the back of solid targets, resulting in the generation of quasi-monoenergetic positron beams.

In experiments, significant progress has been made in electron plasma wakefield acceleration. Due to a lack of sources, there has been relatively less experimental research on positron acceleration, mainly focusing on confirmatory experiments. For instance, the positron-driven hollow plasma wakefield structure conducted at SLAC achieved a maximum acceleration gradient on the order of hundreds of $\mathrm{MV/m}$, which is about 2-3 orders of magnitude lower compared to electron-driven scenarios. Therefore, there is much work to be done in positron acceleration experiments. 
\begin{table}[h]
\centering
\begin{tabular}{p{2.8cm}p{3cm}p{3.5cm}p{5cm}}
\hline
Experiments & Driving bunch & Bunch energy/charge  & Bunch size \tabularnewline
\hline
SLAC\cite{Blumenfeld} & electron bunch & 42 GeV/3.2 nC & $\sigma_z<20\mu m$
\tabularnewline
SLAC\cite{Blue}  & positron bunch & 28.5 GeV/$1.2\times 10^{10}$ & $\sigma_z=2.4ps$
\tabularnewline
FACET\cite{Litos}  & electron bunch & 23 GeV/3.2 nC & $\sigma_z=\sigma_{x,y}=20-30\mu m$
\tabularnewline
FACET\cite{Litos}  & positron bunch & 20.35 GeV/2.2 nC & $\sigma_z=30-50\mu m, \sigma_{x,y}<100\mu m$
\tabularnewline
AWAKE\cite{Bracco} & proton bunch & 400 GeV/$3\times 10^{11}$ &$\sigma_z=120mm,\sigma_{x,y}=200\mu m$
\tabularnewline
AWAKE\cite{Corde} & proton bunch & 300 MeV/3 nC &$\sigma_z=3\mu m$
\tabularnewline
BELLA-LPA \cite{Gonsalves} & laser & 0.85 PW & \tabularnewline
\hline
\end{tabular}
\caption{The relevant parameters of the provided drive beam in the experiment.}
\end{table}

\begin{figure}
	\centering
	\includegraphics[width=\columnwidth]{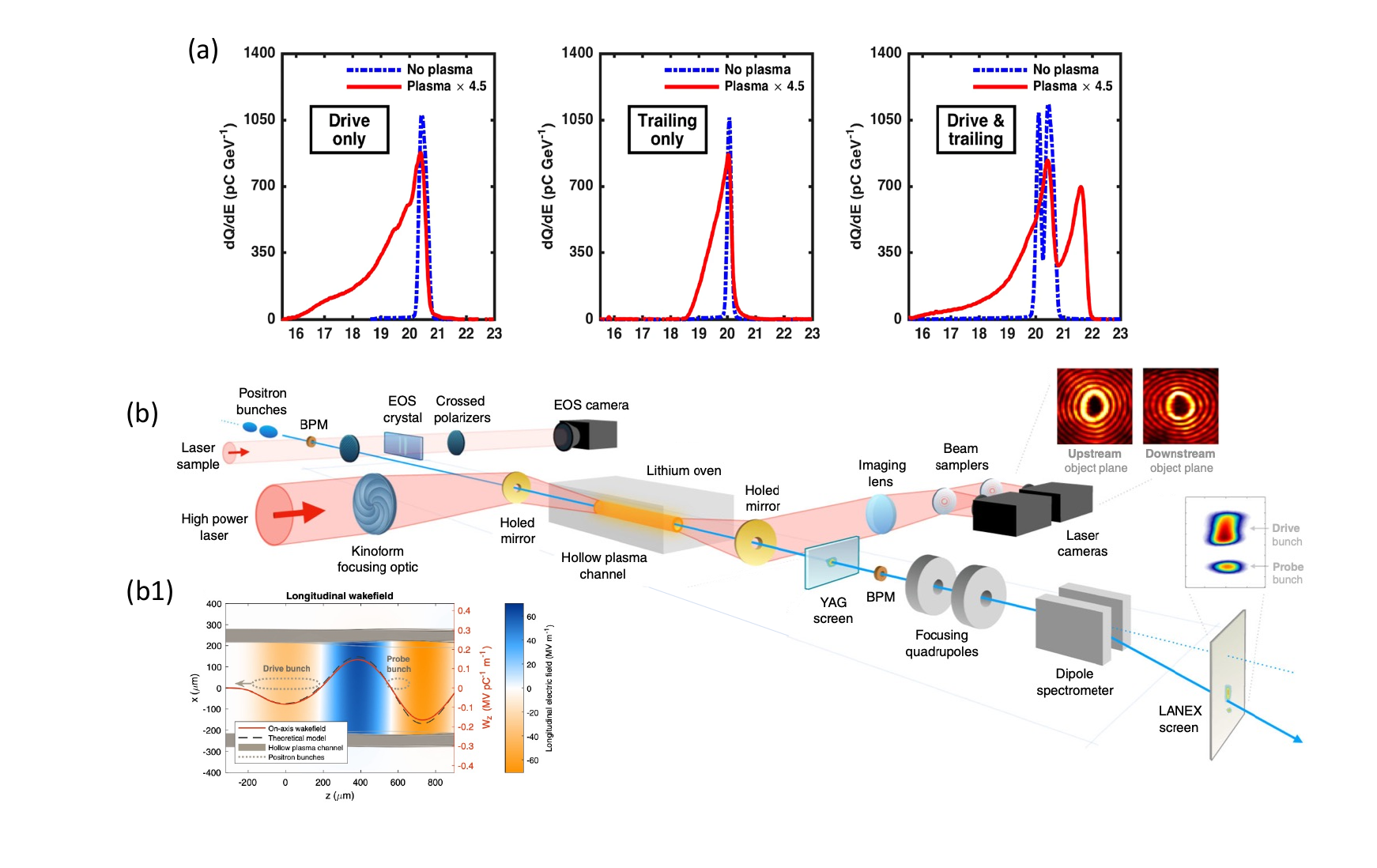}
	\caption{({\bf a}) shows the acceleration effects of the two positron bunches. The energy spectrometer measured with and without plasma (ionizing laser off), the energy gain of the trailing bunch is about 1.45 $\mathrm{GeV}$. From A. Doche et al. \cite{Doche}. ({\bf b}) Experimental setup: the two bunches originated from one bunch with the energy of 20.35 $\mathrm{GeV}$ and the total charge of 0.51 $\mathrm{nC}$, the beam sizes of $\sigma_x=35 \mu m$ and $\sigma_y=25\mu m$. The experiments involve the process of laser ionization to form a hollow plasma channel, as well as the measurement of the field size, the ({\bf b1}) shows the size of the acceleration field and the positions of the drive beam and witness positron beam. From C. A. Lindstr$\varnothing$m et al.\cite{Linds}. }
	\label{Figure4}
\end{figure}

\subsection{Conclusion}
In summary, the proposal of plasma accelerators has overcome the limitations of traditional accelerators, allowing accelerated particles to achieve high energy while promoting the development of compact accelerators. This presents a significant advantage for the construction and cost-effectiveness of future positron-electron colliders. Plasma wakefield accelerators, driven by intense lasers or charged particle beams, generate plasma wakefields that accelerate charged particle beams in the proper phase. Substantial progress has been made in electron acceleration, with energy gains reaching up to 8 $\mathrm{GeV}$. However, there is still a need for improvement in positron beam acceleration. Currently, numerous methods for positron acceleration have been proposed based on Particle-in-Cell (PIC) simulations, but experimental challenges persist. We hope to find methods that are more experimentally feasible to obtain high-energy positron sources, reaching energies in the range of GeV or even higher, up to hundreds of GeV.

\section{acknowledgment}
This work is supported in part by National Natural Science Foundation
of China (11655003); Innovation Project of IHEP (542017IHEPZZBS11820,
542018IHEPZZBS12427); the CAS Center for Excellence in Particle Physics
(CCEPP); IHEP Innovation Grant (Y4545170Y2); Chinese Academy of Science
Focused Science Grant (QYZDY-SSW-SLH002); Chinese Academy of Science
Special Grant for Large Scientific Projects (113111KYSB20170005);
National 1000 Talents Program of China; the National Key Research
and Development Program of China (No.2018YFA0404300).

\section{data availability}
The data that support the findings of this study are available from the corresponding author upon reasonable request.

\end{spacing}

\begin{thebibliography}{}
\vspace{3mm}
\bibitem{Simakov}Simakov, Evgenya I. ,  V. A. Dolgashev, and  S. G. Tantawi, "Advances in high gradient normal conducting accelerator structures," {\it Nuclear Instruments and Methods in Physics Research\/} 907, 221-230(2018).
\bibitem{Lilje}Lilje, L., "High-gradient superconducting radiofrequency cavities for particle acceleration." (2006).
\bibitem{Tajima}Tajima, T. , and  J. Dawson, "Laser Electron Accelerator," {\it Physical Review Letters\/} 43, 267-270(1979).
\bibitem{Gschwendtner}E. Gschwendtner, P. Muggli, "Plasma wakefield accelerators," {\it Nature Reviews Physics\/} 1, 246-248 (2019).
\bibitem{Dunne}Dunne, Mike, "Laser-driven particle accelerators," {\it Science\/} 312.5772(2006):p.374-376.
\bibitem{Bingham}Bingham, R ,  J. T. Mendonca , and  P. K. Shukla, "Plasma based charged-particle accelerators," {\it Plasma physics and controlled fusion\/} 1, 46(2004).
\bibitem{Griffiths}Griffiths, David, "Introduction to Elementary Particles," {\it Academic Press\/} (2008).
\bibitem{Guessoum}Guessoum, Nidhal, "Positron astrophysics and areas of relation to low-energy positron physics," {\it European Physical Journal D\/} 68, 1-6 (2014).
\bibitem{Marchut}L. Marchut, C. J. Mcmahon, "Electron and positron spectroscopies in materials science and engineering," {\it Electron Positron Spectroscopies in Materials Science Engineering\/}, 1–33 (1979).
\bibitem{Ciarmiello}Ciarmiello, Andrea, and L. Mansi, "PET-CT and PET-MRI in neurology: SWOT analysis applied to hybrid imaging."(2016).
\bibitem{Hogstrom}K. R. Hogstrom, P. R. Almond, "Review of electron beam therapy physics," {\it Physics in Medicine, Biology\/} 51, R455 (2006).
\bibitem{ILC}Baer, Howard, et al. "The International Linear Collider Technical Design Report - Volume 2: Physics," {\it Physics\/} 58.7-9, 2901 - 2903(2013).
\bibitem{clic}R. Tomás. "Overview of the Compact Linear Collider," {\it Physical Review Special Topics Accelerators and Beams\/} (2010).
\bibitem{cepc}CEPC Study Group, "CEPC conceptual design report: Volume 2-physics and detector," arXiv preprint arXiv:1811.10545(2018).
\bibitem{Bracco}Bracco, C. , et al. "AWAKE: A Proton-Driven Plasma Wakefield Acceleration Experiment at CERN," {\it Nuclear and Particle Physics Proceedings\/} 273-275.4, 175-180(2016).
\bibitem{Hogan}Hogan, M J , et al. "Plasma wakefield acceleration experiments at FACET," {\it New Journal of Physics\/} 12.5, 055030(2010).
\bibitem{Leemans}Leemans, W. P. , et al. "The BErkeley Lab Laser Accelerator (BELLA): A 10 GeV Laser Plasma Accelerator," {\it American Institute of PhysicsAIP\/} (2010).
\bibitem{Mangles}Mangles, S., Murphy, C., Najmudin, Z. et al. "Monoenergetic beams of relativistic electrons from intense laser–plasma interactions," {\it Nature\/} 431, 535–538 (2004). https://doi.org/10.1038/nature02939
\bibitem{Leemans-W-P}Leemans, W. P. , et al, "Multi-GeV Electron Beams from Capillary-Discharge-Guided Subpetawatt Laser Pulses in the Self-Trapping Regime," {\it Physical Review Letters\/} 113.24(2014):245002.
\bibitem{Faure}J. Faure, Y. Glinec, A. Pukhov et al, "A laser–plasma accelerator producing monoenergetic electron beams," {\it Nature\/} 431, 541-544 (2004).
\bibitem{Gonsalves}Gonsalves, Anthony . "Petawatt laser guiding and electron beam acceleration to 8 GeV in laser-heated capillary discharge waveguides," {\it American Physical Society\/} (2019).
\bibitem{Lotov}Lotov, and  V. K. . "Acceleration of positrons by electron beam-driven wakefields in a plasma," {\it Physics of Plasmas\/} 14.2(2007):693-88.
\bibitem{Alejo}Alejo, Aaron ,  R. Walczak , and  G. Sarri . "Laser-driven high-quality positron sources as possible injectors for plasma-based accelerators," {\it Scientific Reports\/} 1(2019).
\bibitem{Chen}Chen, Hui et al. “Relativistic quasimonoenergetic positron jets from intense laser-solid interactions,” {\it Physical review letters\/} vol. 105,1 (2010): 015003. doi:10.1103/PhysRevLett.105.015003
\bibitem{Diederichs}Diederichs, S. and Benedetti, C. and Esarey, E. and Osterhoff, J. and Schroeder, C. B.. "High-quality positron acceleration in beam-driven plasma accelerators," {\it Phys. Rev. Accel. Beams}23, 121301(2020).
\bibitem{Wang}Wang, X. , et al. "Positron Injection and Acceleration on the Wake Driven by an Electron Beam in a Foil-and-Gas Plasma." {\it Physical Review Letters\/} 101.12, 124801(2008).
\bibitem{Lee}Lee S, Katsouleas T, Hemker RG, Dodd ES, Mori WB. "Plasma-wakefield acceleration of a positron beam". {\it Phys Rev E Stat Nonlin Soft Matter Phys\/} 64, 045501(2001). 
\bibitem{Blue}Blue, B E. , et al. "Plasma-wakefield acceleration of an intense positron beam," {\it Physical review letters\/} 90.21, 214801(2003).
\bibitem{Corde}Corde S, Adli E, Allen JM, et al. "Multi-gigaelectronvolt acceleration of positrons in a self-loaded plasma wakefield," {\it Nature\/} 524(7566):442-445 (2015). doi:10.1038/nature14890.
\bibitem{L-L-Yu}L.-L. Yu, C. B. Schroeder, F.-Y. Li, C. Benedetti, M. Chen, S.-M. Weng, Z.-M. Sheng, E. Esarey. "Control of focusing fields for positron acceleration in nonlinear plasma wakes using multiple laser modes," {\it Physics of Plasmas} 21 (12): 120702(2014).
\bibitem{Vieira} Vieira, Jorge, and J. T. Mendonça. "Nonlinear laser driven donut wakefields for positron and electron acceleration," {\it Physical Review Letters\/} 112.21, 215001 (2014).
\bibitem{Thales} Silva, Thales and Vieira, Jorge. "Positron acceleration in plasma waves driven by non-neutral fireball beams," {\it Phys. Rev. Accel. Beams} 26, 091301(2023).
\bibitem{Jain} Jain N, Antonsen TM Jr, Palastro JP. "Positron Acceleration by Plasma Wakefields Driven by a Hollow Electron Beam," {\it Physical Review Letters\/}115(19):195001(2015). doi:10.1103/PhysRevLett.115.195001
\bibitem{Kimura}Kimura, W. D. , et al. "Hollow plasma channel for positron plasma wakefield acceleration," {\it Physical Review Special Topics - Accelerators and Beams\/} (2011).
\bibitem{Silva}Silva, T. and Amorim, L. D. and Downer, M. C. and Hogan, M. J. and Yakimenko, V. and Zgadzaj, R. and Vieira, J. "Stable Positron Acceleration in Thin, Warm, Hollow Plasma Channels," {\it Physical review letters\/} 127.6(2021).doi:10.1103/PhysRevLett.127.104801.
\bibitem{Yi}Yi, L., Shen, B., Ji, L. et al. "Positron acceleration in a hollow plasma channel up to TeV regime," {\it Scientific Reports\/} 4, 4171 (2014). https://doi.org/10.1038/srep04171
\bibitem{Wei}Zhou, Shiyu and Hua, Jianfei and An, Weiming and Mori, Warren B. and Joshi, Chan and Gao, Jie and Lu, Wei. "High Efficiency Uniform Wakefield Acceleration of a Positron Beam Using Stable Asymmetric Mode in a Hollow Channel Plasma," {\it Physical Review Letters\/} 127, 5 (2021). https://link.aps.org/doi/10.1103/PhysRevLett.127.174801
\bibitem{Xu}Xu, Z., Yi, L., Shen, B. et al. "Driving positron beam acceleration with coherent transition radiation," {\it Communications Physics\/} 3, 191 (2020). https://doi.org/10.1038/s42005-020-00471-6
\bibitem{Yan-Ting}Zhao, J., Hu, YT., Lu, Y. et al. "All-optical quasi-monoenergetic GeV positron bunch generation by twisted laser fields," {\it Communication Physics} 5, 15 (2022).
\bibitem{Terzani}Terzani, Davide and Benedetti, Carlo and Bulanov, Stepan S. and Schroeder, Carl B. and Esarey, Eric. "Compact, all-optical positron production and collection scheme," {\it Phys. Rev. Accel. Beams} 26, 113401(2023).
\bibitem{Xu2}Xu, Zhangli and Shen, Baifei and Si, Meiyu and Huang, Yongsheng, "Positron acceleration by terahertz wave and electron beam in plasma channel,". {\it New Journal of Physics\/} 25(6), 063013(2023).
\bibitem{Si}Meiyu Si, Yongsheng Huang, Manqi Ruan, Baifei Shen, Zhangli Xu, Tongpu Yu, Xiongfei Wang, and Yuan Chen, "Relativistic-guided stable mode of few-cycle 20 $\mathrm{\mu m}$ level infrared radiation," {\it Optics. Express\/} 31, 40202-40209 (2023).
\bibitem{Si2} Si Meiyu, Huang Yongsheng, Ruan Manqi, Shen Baifei, Xu Zhangli, Yu Tongpu, Wang Xiongfei, Chen Yuan, "Novel positron acceleration by continuous coherent mid-infrared radiation in a micro-tube," {\it arXiv preprint\/} arXiv:2302.12418 (2023).
\bibitem{Jie-zhao}Zhao Jie, Li QN, Hu YT, Zhang H, Cao Y, Sha R, Shao FQ, Yu TP. et al, "Terahertz-driven positron acceleration assisted by ultra-intense lasers," {\it Optics express} vol. 31,14 (2023): 
\bibitem{Zhao}Zhao, Jie, Yan-Ting Hu, Hao Zhang, Yu Lu, Li-Xiang Hu, Fu-Qiu Shao, and Tong-Pu Yu. "Multistage Positron Acceleration by an Electron Beam-Driven Strong Terahertz Radiation," {\it Photonics\/} 10, no. 4: 364 (2023).
\bibitem{Chernyaev}Chernyaev, A.P., Varzar, S.M, "Particle accelerators in modern world," {\it Physics of Atomic  Nuclei} 77, 1203–1215 (2014). 
\bibitem{Conard}Conard, E. Milo. "High intensity accelerator for a wide range of applications," {\it Nuclear Instruments and Methods in Physics Research Section A: Accelerators, Spectrometers, Detectors and Associated Equipment} 353.1-3, 1-5(1994).
\bibitem{Amaldi}Amaldi, Ugo. "Accelerators for Medical Applications." (1996).
\bibitem{Wieszczycka}Wieszczycka, Wioletta, and Waldemar H. Scharf. Warsaw University of Technology, Poland, "Proton Therapy Accelerators”.(2001).
\bibitem{Bhandari}R.K. Bhandari, Malay Kanti Dey, "Applications of Accelerator Technology and its Relevance to Nuclear Technology," {\it Energy Procedia} 7, 577-588(2011).
\bibitem{offe}Ioffe, Boris L. . "The origin of mass and experiments on future high-energy accelerators," {\it Physics-Uspekhi} (2007).
\bibitem{Cline}Cline, D. B. , et al. "A compact hard X-ray source for medical imaging and biomolecular studies," {\it Nuclear Instruments  Methods in Physics Research} 99.1-4, 821-823(1995).
\bibitem{Sidky}E. Sidky, Emil Y. , L. Yu , and X. Pan . "Application of expectation maximization to x-ray spectrum estimation for medical accelerators from transmission data," {\it Physics of Medical Imaging} pt.2 (2004).
\bibitem{Selim}Selim, et al. "Bremsstrahlung-induced highly penetrating probes for nondestructive assay and defect analysis," {\it Nuclear Instruments Methods in Physics Research Section A} (2002).
\bibitem{Gourlay}Gourlay, Stephen and Raubenheimer, Tor and Shiltsev, Vladimir, "Challenges of Future Accelerators for Particle Physics Research," {\it Frontiers in Physics} (2022).
\bibitem{lhc}Benedikt, Michael , et al. "LHC Design Report," CERN (2004).
\bibitem{Nakajima}Nakajima, Kazuhisa. "Laser accelerator developments for future high-energy accelerators," {\it Nuclear Instruments and Methods in Physics Research Section A: Accelerators, Spectrometers, Detectors and Associated Equipment} 410. 3, 514-519(1998).
\bibitem{Hooker}Hooker, S. "Developments in laser-driven plasma accelerators," {\it Nature Photon} 7, 775–782 (2013). 
\bibitem{Rosenzweig}Rosenzweig, J. B. , et al. "Acceleration and focusing of electrons in two-dimensional nonlinear plasma wake fields," {\it Physical Review A} (1991).
\bibitem{Litos(2014)}Litos, M., Adli, E., An, W. et al. "High-efficiency acceleration of an electron beam in a plasma wakefield accelerator," {\it Nature }515, 92–95 (2014). 
\bibitem{Huang}Huang, C. , et al. "Simulation of a 50GeV PWFA Stage," {\it American Institute of Physics} (2004).
\bibitem{Lu}Lu, Wei , et al. "Design and simulation of a single 100GeV stage Laser Wakefield Accelerator," {\it APS Meeting Abstracts} (2007).
\bibitem{Kurz}Kurz, T., Heinemann, T., Gilljohann, M.F. et al. "Demonstration of a compact plasma accelerator powered by laser-accelerated electron beams," {\it Nat Commun} 12, 2895 (2021). 
\bibitem{Malka}Malka, V. "Plasma Wake Accelerators: Introduction and Historical Overview," {\it CERN Yellow Reports}, 1 (2016).
\bibitem{Schwoerer}Schwoerer, H. and Liesfeld, B. and Schlenvoigt, H.-P. and Amthor, K.-U. and Sauerbrey, R. "Thomson-Backscattered X Rays From Laser-Accelerated Electrons," {\it Phys. Rev. Lett.} 96, 4(2006).
\bibitem{Rousse}Rousse, A. et al. "Production of a keV X-ray beam from synchrotron radiation in relativistic laser–plasma interaction," {\it Phys. Rev. Lett}. 93, 135005 (2004).
\bibitem{Hartemann}Hartemann, F. V. et al. "Compton scattering X-ray sources driven by laser wakefield acceleration," {\it Phys. Rev. ST Accel Beams} 10, 011301 (2007).
\bibitem{Schlenvoigt}Schlenvoigt, HP., Haupt, K., Debus, A. et al. "A compact synchrotron radiation source driven by a laser-plasma wakefield accelerator," {\it Nature Phys} 4, 130–133 (2008). 
\bibitem{Gruener}Gruener, F. et al. "Design considerations for table top laser based VUV and X-ray free electron lasers," {\it Appl. Phys. B} 86, 431–435 (2007).
\bibitem{Malka(2008)}Malka, V., Faure, J., Gauduel, Y. et al. "Principles and applications of compact laser–plasma accelerators," {\it Nature Phys} 4, 447–453 (2008). https://doi.org/10.1038/nphys966
\bibitem{Kostyukov}I. Kostyukov, S. Kiselev, A. Pukhov. "X-ray generation in an ion channel," {\it Phys. Plasmas} 1 December 2003; 10 (12): 4818–4828. https://doi.org/10.1063/1.1624605
\bibitem{Pak}Pak, T., Rezaei-Pandari, M., Kim, S.B. et al. "Multi-millijoule terahertz emission from laser-wakefield-accelerated electrons," {\it  Light Sci Appl }12, 37 (2023). 
\bibitem{Hugenschmidt}C. Hugenschmidt, K. Schreckenbach, M. Stadlbauer, B. Straßer. "Low-energy positrons of high intensity at the new positron beam facility NEPOMUC," {\it Nuclear Instruments and Methods in Physics Research Section A: Accelerators, Spectrometers, Detectors and Associated Equipment}554, 384-391(2005).
\bibitem{Chaikovska}Chaikovska, I., Chehab, R., Kubytskyi, V., Ogur, S., Ushakov, A., Variola, A., Sievers, P., Musumeci, P., Bandiera, L., Enomoto, Y., Hogan, M. J., and Martyshkin, P.. "Positron sources: from conventional to advanced accelerator concepts-based colliders," {\it Journal of Instrumentation} 17(2022).
\bibitem{Golge}Golge, Serkan and Branislav Vlahovic. “REVIEWOF LOW-ENERGY POSITRON BEAM FACILITIES.” (2012).
\bibitem{Y-H-Chen-2018}Y. H. Chen et al., "GeV positron production via the multi-photon Breit-Wheeler process in aligned crystals," {\it Nuclear Instruments and Methods in Physics Research Section B: Beam Interactions with Materials and Atoms} 417, 16-19 (2018).
\bibitem{FCC}FCC collaboration, "FCC-ee: The Lepton Collider: Future Circular Collider Conceptual Design Report," {\it Eur. Phys. J. ST} 228(2019).
\bibitem{Nanni2015}Nanni, E., Huang, W., Hong, KH. et al, "Terahertz-driven linear electron acceleration." {\it Nature Communications} 6, 8486 (2015). https://doi.org/10.1038/ncomms9486
\bibitem{Hibberd2020}Hibberd, M.T., Healy, A.L., Lake, D.S. et al, "Acceleration of relativistic beams using laser-generated terahertz pulses." {\it Nature Photonics} 14, 755–759 (2020). https://doi.org/10.1038/s41566-020-0674-1.
\bibitem{Xu2021}Xu, H., Yan, L., Du, Y. et al, "Cascaded high-gradient terahertz-driven acceleration of relativistic electron beams." {\it Nature Photonics} 15, 426–430 (2021). https://doi.org/10.1038/s41566-021-00779-x.
\bibitem{Yu2023}Yu, XQ., Zeng, YS., Song, LW. et al, "Megaelectronvolt electron acceleration driven by terahertz surface waves." {\it Nature Photonics} 17, 957–963 (2023). https://doi.org/10.1038/s41566-023-01251-8.
\bibitem{Martinez}Martinez, Bertrand and Barbosa, Bernardo and Vranic, Marija. "Creation and direct laser acceleration of positrons in a single stage," {\it Phys. Rev. Accel. Beams} 26, 7(2023).
\bibitem{Yu} Cao, Y and Hu, L X and Zou, D B and Yang, X H and Hu, Y T and Zhao, J and Lu, Y and Yin, Y and Shao, F Q and Yu, T P. "Collimation, compression and acceleration of isotropic hot positrons by an intense vortex laser," {\it New Journal of Physics} 25(2023).
\bibitem{Gordon}D. F. Gordon, B. Hafizi, R. F. Hubbard, et al., "Asymmetric self-phase modulation and compression of short laser
pulses in plasma channels," {\it Phys. Rev. Lett} 90(21), 215001 (2003).
\bibitem{Nie}Z. Nie, C.-H. Pai, J. Hua, et al., "Relativistic single-cycle tunable infrared pulses generated from a tailored plasma
density structure," {\it Nat. Photonics} 12(8), 489–494 (2018).
\bibitem{Zhu}X.-L. Zhu, S.-M. Weng, M. Chen, et al., "Efficient generation of relativistic near-single-cycle mid-infrared pulses in
plasmas," {\it Light: Sci. Appl} 9(1), 46 (2020).
\bibitem{Nie2} Z. Nie, C.-H. Pai, J. Zhang, et al., "Photon deceleration in plasma wakes generates single-cycle relativistic tunable
infrared pulses," {\it Nature Communication} 11(1), 2787 (2020).
\bibitem{Joshi(2018)} C Joshi and E Adli and W An and C E Clayton and S Corde and S Gessner and M J Hogan et al. "Plasma wakefield acceleration experiments at FACET II," {\it Plasma Physics and Controlled Fusion} 60, 034001(2018).
\bibitem{Blumenfeld} Blumenfeld, I., Clayton, C., Decker, FJ. et al. "Energy doubling of 42 GeV electrons in a metre-scale plasma wakefield accelerator," {\it Nature} 445, 741–744 (2007). 
\bibitem{Litos}Litos, Michael and Adli, E and Allen, J and An, Weiming and Clarke, C and Corde, Sébastien and Clayton, et al. "9 GeV Energy Gain in a Beam-Driven Plasma Wakefield Accelerator," {\it Plasma Physics and Controlled Fusion} 58 (2015).
\bibitem{Steinke}Steinke, S., van Tilborg, J., Benedetti, C. et al. "Multistage coupling of independent laser-plasma accelerators," {\it Nature} 530, 190–193 (2016).
\bibitem{Doche}Doche, A., Beekman, C., Corde, S. et al. "Acceleration of a trailing positron bunch in a plasma wakefield accelerator," {\it Sci Rep} 7, 14180 (2017). 
\bibitem{Gessner}Gessner, S., Adli, E., Allen, J. et al. "Demonstration of a positron beam-driven hollow channel plasma wakefield accelerator," {\it Nature  Communication} 7, 11785 (2016). 
\bibitem{Linds}Lindstr$\varnothing$m, C. A. and Adli, E. and Allen, J. M. and An, W. et al. "Measurement of Transverse Wakefields Induced by a Misaligned Positron Bunch in a Hollow Channel Plasma Accelerator," {\it Phys. Rev. Lett.} 120, 124802(2018).
\bibitem{Yonghong}Yonghong Yan and Yuchi Wu and Jia Chen and Minghai Yu and Kegong Dong and Yuqiu Gu. "Positron acceleration by sheath field in ultra-intense laser–solid interactions," {\it Plasma Physics and Controlled Fusion} 59, 045015(2017).



\end{thebibliography}
\end{document}